\documentclass[aps,preprint]{revtex4}%
\usepackage{amsfonts}
\usepackage{amsmath}
\usepackage{amssymb}
\usepackage{graphicx}%
\begin{document}
\title{Above-threshold ionization photoelectron spectrum from quantum trajectory}
\author{X. Y. Lai$^{1,2}$}
\author{Q. Y. Cai$^{1}$}
\email{qycai@wipm.ac.cn}
\author{M. S. Zhan$^{1,3}$}
\affiliation{$^{1}$State Key Laboratory of Magnetic Resonances and
Atomic and Molecular Physics, Wuhan Institute of Physics and
Mathematics, The Chinese Academy of Sciences, Wuhan 430071, P.R.
China} \affiliation{$^{2}$Graduation University of Chinese Academy
of Sciences, Beijing 100081, P.R. China} \affiliation{$^{3}$Center
for Cold Atom Physics, The Chinese Academy of Sciences, Wuhan
430071, P.R. China}

\begin{abstract}
Many nonlinear quantum phenomena of intense laser-atom physics can
be intuitively explained with the concept of trajectory. In this
paper, Bohmian mechanics (BM) is introduced to study a multiphoton
process of atoms interacting with the intense laser field:
above-threshold ionization (ATI). Quantum trajectory of an atomic
electron in intense laser field is obtained from the Bohm-Newton
equation first and then the energy of the photoelectron is gained
from its trajectory. With energies of an ensemble of photoelectrons,
we obtain the ATI spectrum which is consistent with the previous
theoretical and experimental results. Comparing BM with the
classical trajectory Monte-Carlo method, we conclude that quantum
potential may play a key role to reproduce the spectrum of ATI. Our
work may present a new approach to understanding quantum phenomena
in intense laser-atom physics with the image of trajectory.

\end{abstract}

\maketitle

\section{Introduction}
In intense laser field (ILF), an atom may absorb multiple photons,
more than that required for ionization, and then it will eject a
high energy photoelectron. Such a nonlinear phenomenon of
multiphoton process is called above-threshold ionization (ATI)
\cite{Burnett,Delone,Mittleman1}. In general, the photoelectron
spectrum of ATI consists of multiple peaks, separated by one photon
energy \cite{Martin,Agostini}. Many theoretical methods have been
developed to study this multiphoton phenomenon, such as solving the
time-dependent Schr\"{o}dinger equation \cite{Javanainen,zhou}, or
the semiclassical trajectory methods including two-step model
\cite{Corkum} and Feynman's path-integral approach in the strong
field approximation \cite{Salieres,Salieres2,eden}.

Bohmian mechanics (BM) \cite{Bohm,Holland,Nikolic}, or called
quantum trajectory method \cite{rew}, is an alternative to quantum
mechanics. It has been successfully used to study some fundamental
quantum phenomena such as tunneling \cite{joh,cd} and the
two-slit experiment \cite{cp}. Recently, BM has been extensively
applied to study many novel quantum processes in physics and
chemistry, such as atom-surface physics \cite{ass3,ass}, electron
transport in mesoscopic systems \cite{xo}, photodissociation of NOCl
and NO$_{2}$ \cite{bkd}, and the chemical reactions \cite{rew2}. It
has also been regarded as a resultful approach to studying chaos
\cite{us,mhp,daw,ce} and decoherence \cite{ass2}.

In this paper, BM is introduced to intense laser-atom physics to
study the multiphoton phenomenon of ATI. The trajectory of each
atomic electron in an ensemble can be deterministically obtained
from the Bohm-Newton equation which is a subtle transformation of
the Schr\"{o}dinger equation and equal to it statistically
\cite{Bohm}. Then the energy of an ionized electron, or called
photoelectron, can be calculated from its trajectory. By rearranging
all photoelectrons according to their energies, we gain a
photoelectron spectrum of ATI which is consistent with the previous
theoretical and experimental results. Finally, we briefly discuss
the classical trajectory Monte-Carlo method (CTMCM)
and conclude that a term in Bohm-Newton equation, called quantum
potential, may play a key role to precisely reproduce the spectrum
of ATI.

This paper is organized as follow: We will briefly introduce BM
first. Then we show the Hamiltonian for the hydrogen atom in ILF.
The numerical solution of the time-dependent Schr\"{o}dinger
equation and details of BM for ATI together with the photoelectron
spectrum are given. Next we compare our result with the previous
theoretical and experimental results of ATI. Finally, we compare BM
with CTMCM and then conclude.

\section{The quantum trajectories formalism of Bohmian mechanics}
Bohm-Newton equation comes from a subtle transformation of the
time-dependent Schr\"{o}dinger equation \cite{Bohm,Holland}. A wave
function $\psi$ can be expressed as
$\psi(x,t)=R(x,t)e^{iS(x,t)/\hbar}$, where $R$ and $S$ are real
functions. Inserting $\psi(x,t)$ into the time-dependent
Schr\"{o}dinger equation, we obtain two equations by separating the
time-dependent Schr\"{o}dinger equation into real and imaginary
parts. The real part gives

\begin{equation}
\frac{\partial S}{\partial t}+\frac{\left(  \nabla S\right)
^{2}}{2m}+V+Q=0
\label{qp}%
\end{equation}
and the imaginary part has the form%

\begin{equation}
\frac{\partial\rho}{\partial t}+\nabla(\rho v)=0,
\end{equation}
where $Q(x,t)=-\frac{\hbar^{2}}{2m}\frac{\nabla^{2}R}{R}$, $\rho
(x,t)=R^{2}(x,t)$ and $v=\nabla S(x,t)/m$. We find that equation (2)
looks like the classical continuity equation, and equation (1) can
be reduced to the classical Hamilton-Jacobi equation if the term
$Q(x,t)$ was ignored. In BM, $Q(x,t)$ is usually called quantum
potential and plays a crucial role for the appearance of quantum
phenomena \cite{Bohm,Holland}. From the standpoint of classical
mechanics, a Bohm-Newton equation of motion for a Bohmian particle
can be constructed:
\begin{equation}
md^{2} x/dt^{2}=-\nabla(V+Q). \label{bn}%
\end{equation}
In fact, according to the definition above, a much simpler equation
of motion can be used to obtain quantum trajectory instead of
equation (\ref{bn}):
\begin{equation}
dx/dt=\nabla S(x,t)/m\,. \label{dy}%
\end{equation}
Usually, we first solve the time-dependent Schr\"{o}dinger equation
to obtain $\psi(x,t)$, and hence $S(x,t)$. Secondly we integrate
equation (\ref{dy}) to obtain the quantum trajectory of the
particle.

Statistically, Bohm-Newton equation is completely equal to
Schr\"{o}dinger equation. All quantum phenomena described by
Schr\"{o}dinger equation can be reproduced by Bohm-Newton equation,
in principle, with a properly initial ensemble distribution. In
particular, Bohm-Newton equation can help us to understand quantum
world intuitively from the view of trajectory. In semiclassical
method, the multiphoton phenomenon of ATI can be intuitively
explained with the concept of electron trajectory. Therefore,
quantum trajectory method is used to study the phenomenon of ATI in
intense laser-atom physics in this paper.

\section{Numerical solution of the time-dependent Schr\"{o}dinger equation}
In this paper, we study the system of hydrogen atom in ILF, using
the one-dimensional model atom of hydrogen
\cite{Javanainen,zhou,Rz,qiao}. The one-dimensional model, or namely
the soft-core model, which has the long-range Coulomb tail
characteristic of real atomic system, can approximate the system of
hydrogen asymptotically. The Hamiltonian of the field-free atom is
$H_{0}(x)=-\frac{1}{2}\frac{d^{2}}{dx^{2}}-\frac{1}{\sqrt{1+x^{2}}}$
and the atom-laser interaction is $H_{t}(x)=-xE(t)$, where $E(t)$ is
the laser field profile (atomic units are used throughout). Thus the
time-dependent Schr\"{o}dinger equation can be obtained
$i\frac{\partial\psi(x,t)}{\partial
t}=[H_{0}(x)+H_{t}(x)]\psi(x,t)$. To solve the time-dependent
Schr\"{o}dinger equation, we first need to gain the eigenstates and
eigenvalues of $H_{0}(x)$: $\varphi_{n}(x)$ and $E_{n}$
$(n=1,2,...,N)$, respectively. Here we construct the eigenstates
from the B-spline basis,
$\varphi_{n}(x)=\sum\limits_{j=1}^{M}C_{j}^{n}B_{j}^{k}(x)$ where
$B_{j}^{k}(x)$ is the B-spline basis and $k$ is the order of
B-spline basis \cite{xi}. The range of the variable $x$ is confined
to $(-x_{\max},x_{\max})$. Using the standard diagonalization
procedure, the eigenvalues $E_{n}$ and values $C_{j}^{n}$ are
obtained respectively. Secondly, the time-dependent wavefunction
$\psi(x,t)$ is expressed in terms of the eigenstates as
$\psi(x,t)=\sum\limits_{n}a_{n}(t)\varphi_{n}(x)$, where $a_{n}(t)$
is a time-dependent function. In this work we use Symplectic
Algorithm method \cite{sanz-serna} to solve the time-dependent
Schr\"{o}dinger equation to obtain the coefficient $a_{n}(t)$, and
hence the wavefunction $\psi(x,t)$.

In the present paper, the laser field profile is $E(t)= \left\{
\begin{array}{ll}
E_{0}\sin^{2}(\frac{\pi t}{6T})\sin(\omega t), & {0\leq t\leq 3T}\\
E_{0}\sin(\omega t), & { t> 3T}
\end{array}
\right.  $,
 where
$T=2\pi/\omega$, and $E_{0}$ and $\omega$ are the amplitude and
angular frequency of the electric field in the laser pulse,
respectively. Here $E_{0}=0.1$ a.u. and $\omega=0.148$ a.u. And we
take: $k=7,$ $M=1200,$ $N=1028$ and $x_{\max}=600$ a.u., with the
time step of $0.0051824$ a.u. The initial state $\psi(x,0)$ of the
system is the ground state  of the field-free one-dimensional model
atom of hydrogen.

\section{Results}

%
\begin{figure}
\resizebox{0.75\columnwidth}{!}{%
  \includegraphics{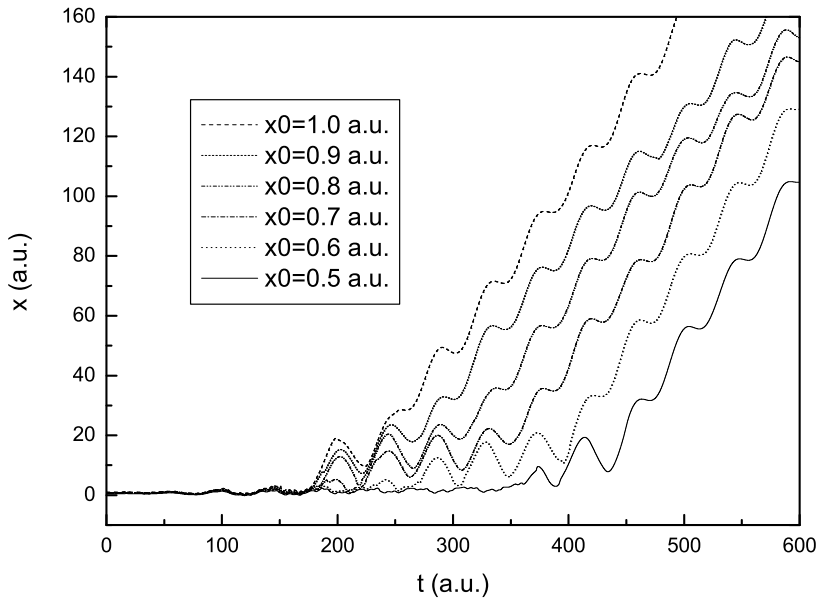}
}
\caption{Quantum trajectories of electrons with different initial
positions as functions of time. The initial positions $x_{0}$ are,
from top to botton, $1.0$ a.u., $0.9$ a.u., $0.8$ a.u., $0.7$ a.u.,
$0.6$ a.u., and $0.5$ a.u.}
\label{fig1}       
\end{figure}

After obtaining the time-dependent wavefunction $\psi(x,t)$ and the
companying $S(x,t)$, we can numerically integrate equation
(\ref{dy}) to get the quantum trajectory of electron with its
initial position $x_{0}$. Figure \ref{fig1} shows the different
electron trajectories with the corresponding initial positions as
functions of time. Let's take the trajectory with the initial
position $x_{0}=0.5$ a.u. as an example: At the first stage before
$t=450$ a.u. about, the electron oscillates around the core, driven
by the laser field. After that, the electron runs away from the
core, following a linear line with small periodic oscillation. This
implies that the electron has been ionized, \emph{i.e.}, a
photoelectron has been ejected from the atom. We have obtained lots
of trajectories of electrons with different initial positions
$x_{0}$
. All of them have the similar characters as explicitly shown in
Figure \ref{fig1}. We then calculate the average kinetic energy
$\left\langle E_{K}\right\rangle $ of each ionized electron in one
period of oscillation from its trajectory. Because of the long-tail
potential of the model atom, the energy of the photoelectron is thus
obtained
\begin{equation}
E=\left\langle E_{K}\right\rangle +V(x)_{atom}, \label{e}%
\end{equation}
where $V(x)_{atom}$ is the long-tail potential of the model atom. In
this way, we have shown the energy of a photoelectron can be
obtained directly from the Bohm-Newton equation. In the traditional
semiclassical trajectory methods, however, the energy of
photoelectron is obtained from a semi-empirical formula
\cite{Burnett}: The total energy of a free electron in the laser
field is the
sum of the ponderomotive energy and the translational energy: $E=E_{p}%
+\frac{1}{2}\left\langle v\right\rangle ^{2}$, where $E_{p}=(E_{0}%
/2\omega)^{2}$ and $\left\langle v\right\rangle $ is the electron
translational velocity in the laser field.

%
\begin{figure}
\resizebox{0.7\columnwidth}{!}{%
  \includegraphics{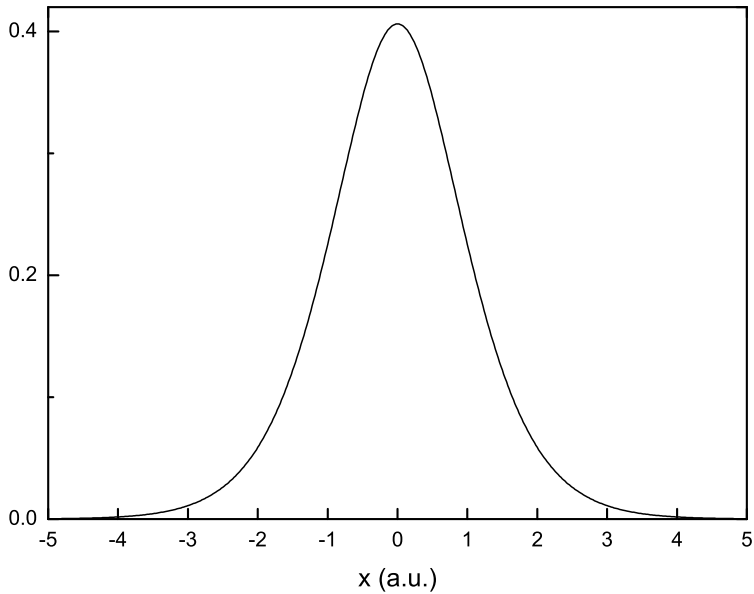}
}
\caption{The initial electron density distribution $|\psi(x,0)|^{2}$
in the range $(-5,5)$.}
\label{fig2}       
\end{figure}
%

%
\begin{figure}
\resizebox{0.75\columnwidth}{!}{%
  \includegraphics{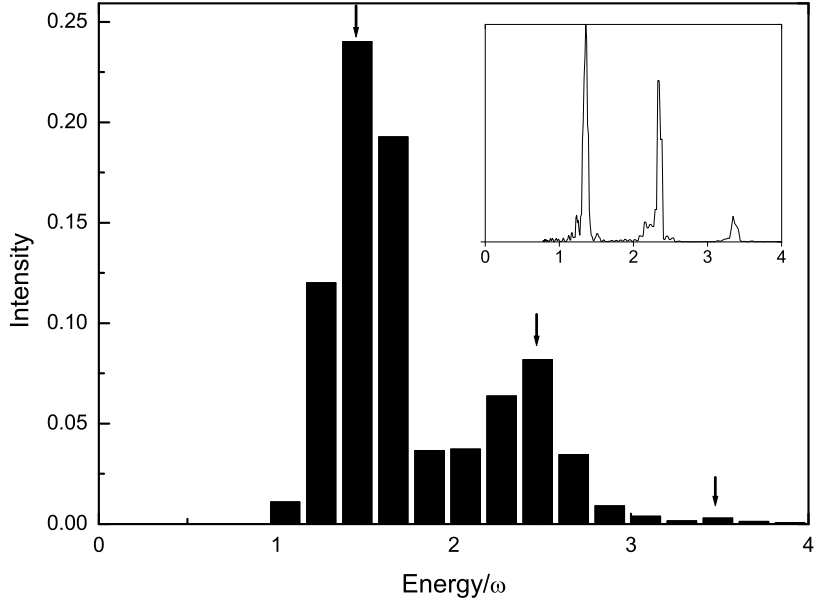}
}
\caption{ATI spectrum with three peaks at about $1.4\omega$,
$2.4\omega$, and $3.4\omega$, respectively. The inset is the ATI
spectrum obtained by solving the time-dependent Schr\"{o}dinger
equation at the time $t=16.25T$, after considering the ponderomotive
after-acceleration of electron in intense laser field.}
\label{fig3}       
\end{figure}

Next, we calculate lots of trajectories of atomic electrons in an
ensemble to gain the energies of each photoelectron, and hence the
photoelectron spectrum of ATI. According to Bohmian theory, the
initial distribution of atomic electrons in an ensemble is
$|\psi(x,0)|^{2}$ \cite{Bohm,Holland}. In this paper, $\psi(x,0)$ is
the ground state  of the field-free one-dimensional model atom of
hydrogen, which is obtained by numerically solving the
time-independent Schr\"{o}dinger equation\textbf{:
$H_{0}\psi=E\psi$}. Figure \ref{fig2} explicitly shows the initial
density distribution of the electrons. We have chosen 8,192 initial
positions distributed in the range $(-5,5)$ with the distribution
density $|\psi(x,0)|^{2}$ (note that $99.9\%$ of atomic electrons in
the ensemble is in the range $(-5,5)$ at the time $t=0$) and gotten
the corresponding trajectories with the propagation time $16T$. Then
we gain the energies of the corresponding photoelectrons by equation
(\ref{e}). At last, by rearranging these photoelectrons according to
their energies, we can gain a photoelectron spectrum of ATI, which
is shown in figure \ref{fig3} with three peaks at about $1.4\omega$,
$2.4\omega$, and $3.4\omega$, respectively. These peaks are
contributed from different quantum trajectories with the
corresponding initial electron positions $x_{0}$. Generally
speaking, the photoelectrons with the initial positions near the
core contribute to the first peak and the photoelectrons with the
initial positions not near the core have the contribution to the
second peak, the third peak, etc. For example, the photoelectron
from the initial position $x_{0}=0.5$ a.u. with the energy
$1.5\omega$ contributes to the first peak, and the one from
$x_{0}=1.8$ a.u. with the energy $2.4\omega$ contributes to the
second peak.

Our result of ATI spectrum from BM is consistent with the previous
experimental and theoretical results. In experiment, the positions
of the peaks in a ATI spectrum are accurately given by the formula
$E=E_{g}^{0}+n\omega$, where $E_{g}^{0}$\ is the ground state energy
of the field-free atom and $n$ is the
number of photons absorbed by the atom \cite{Burnett}. When $E_{g}^{0}+n^{\prime}%
\omega\geq0$ but $E_{g}^{0}+n^{\prime}\omega-E_{p}\leq0$, due to the
AC Stark shift, the channel for $n^{\prime}$-photon absorption is
closed and the corresponding peak in the ATI spectrum is suppressed,
which is called channel closing and has been demonstrated
experimentally \cite{LX}. In our work, $\omega=0.148$ a.u. and the
ground state energy of the model atom is $E_{g}^{0}=-0.6697$ a.u.,
so the peaks should appear at $1.48\omega$, $2.48\omega$, and
$3.48\omega$, respectively. We have also reproduced the ATI spectrum
by solving the time-dependent Schr\"{o}dinger equation
\cite{Javanainen}. After considering the ponderomotive
after-acceleration of electron in ILF \cite{Bucksbaum}, we gain the
ATI spectrum (the inset in figure \ref{fig3}) which agrees with that
from BM. In this way, we have shown that the ATI spectrum we
obtained with BM is consistent with the previous experimental and
theoretical results.

\section{Discussion}

It is interesting that BM can well describe the process of ATI, in
particular after comparing with the application of CTMCM to intense
laser-atom physics. In the past, CTMCM has been used to study the
process of ionization of the hydrogen atom in ILF \cite{cohen}. It
has been definitely pointed out that CTMCM itself cannot give the
correct ionization rate, unless an additional tunneling process was
considered (see figure 1 in \cite{cohen}). In CTMCM, motion of the
electron is dominated by the classical Newton equation. In our
calculation, however, Bohm-Newton equation has been used to gain the
trajectory of the electron. We can thus conclude that the quantum
potential $Q(x,t)$ in equation (\ref{bn}) may play a crucial role to
reproduce the spectrum of ATI in intense laser-atom physics.

\section{Summary}

In summary, we have used BM to describe the process of ATI in
intense laser-atom physics. In our study, the energy of each
photoelectron is obtained from Bohm-Newton equation and the obtained
ATI spectrum is consistent with the previous theoretical and
experimental results. Comparing BM with CTMCM, we conclude that
quantum potential may play a key role to depict the quantum
phenomenon. Therefore, our work may present a new approach to
studying the quantum phenomena in intense laser-atom physics with
the image of trajectory.

\end{document}